\documentclass[12pt]{article}  
\usepackage{amssymb,latexsym}  
\usepackage{amsmath}  
 
\columnwidth\textwidth

\begin{document}

\newtheorem{df}{Definition}  
\newtheorem{thm}{Theorem}  
\newtheorem{lem}{Lemma}  
 
\begin{titlepage}  
 
\noindent  
 
\begin{center}  
{\LARGE Intrinsic properties of quantum systems}  
 
\vspace{1cm}  
 
P. H\'{a}j\'{\i}\v{c}ek \\  
Institute for Theoretical Physics \\  
University of Bern \\  
Sidlerstrasse 5, CH-3012 Bern, Switzerland \\  
hajicek@itp.unibe.ch \\  
\vspace{5mm}  
and \\  
\vspace{5mm}  
J. Tolar\\  
Department of Physics\\  
Faculty of Nuclear Sciences and Physical Engineering\\  
Czech Technical University\\  
B\v{r}ehov\'{a} 7, CZ-11519 Prague, Czech Republic\\  
jiri.tolar@fjfi.cvut.cz  
 
\vspace{1cm}  
 
 
February 2009 \\  
 
 
\vspace*{5mm}  
 
\nopagebreak[4]  
 
\begin{abstract}
A new realist interpretation of quantum mechanics is introduced. Quantum
systems are shown to have two kinds of properties: the usual ones described by
values of quantum observables, which are called extrinsic, and those that can
be attributed to individual quantum systems without violating standard quantum
mechanics, which are called intrinsic. The intrinsic properties are classified
into structural and conditional. A systematic and self-consistent account is
given. Much more statements become meaningful than any version of Copenhagen
interpretation would allow. A new approach to classical properties, quantum
cosmology and measurement problem is suggested. A quantum definition of
classical states is proposed. 
\end{abstract}

\end{center}

\end{titlepage}

\section{Introduction}
Quantum mechanics does not seem to be fully understood even after about eighty
years of its very successful existence and a lot of work is being done on its
interpretation or modification today (e.g., \cite{Griffiths,Nikolic}). The
present paper describes an approach to its conceptual foundation from a new
point of view. 

Accounts of the conceptual structure of quantum mechanics usually start from a
'minimal' interpretative framework on which adherents of different
interpretations can agree. For example in \cite{Isham}, this is referred to as
'pragmatic approach' to quantum theory and a review of it is called 'The Rules
of Quantum Theory', PP. 67-75. In \cite{d'Espagnat}, a similar review is
called 'Rules of Quantum Mechanics', PP. 42-56. From d'Espagnat review, we
leave out Rules 7-10 that concern the so-called ideal measurements (in our
language, ideal measurements are preparations). Isham' Rules 1-4 and
d'Espagnat Rules 1-6 are practically equivalent and where they differ, they
are compatible. We call their logical union {\em standard quantum mechanics}. 

An interpretation of quantum mechanics is defined as those hypotheses that are
added to standard quantum mechanics\footnote{If the hypotheses modify standard
  quantum theory as in, e.g., pilot-wave theory by Bohm and de Broglie
  \cite{Nikolic}, then we do not call the result quantum mechanics.}. For
example, Copenhagen interpretation adds the hypothesis that quantum systems
cannot posses real properties or even that quantum systems do not
exist. Many-world interpretation adds that all values that a measurement can
give really exist in different branches of the world, etc. 

Standard quantum mechanics can be characterised as a set of rules allowing the
computation of probabilities for the outcomes of registrations which follow
specific preparations. The preparations and registrations work only with
classical systems and classical properties, the existence of which is not
denied. The question is, however, left open whether quantum system really
exist and can posses real properties\footnote{In Isham's and d'Espagnat
  account of standard quantum theory, the existence of quantum systems seem to
  be assumed. But this is not what everybody can accept (see the discussion in
  \cite{Peres,ludwig,kraus}: they define quantum systems just as equivalence
  classes of preparations or registrations).}. In dependence of how an
interpretation answers this question, it is classified as realist or
anti-realist. Some criteria are formulated in \cite{Isham}, (P. 68): For an
anti-realist interpretation: 
\begin{quote}
The notion of an individual physical system 'having' or 'possessing' of all
its physical quantities is {\em inappropriate} in the context of quantum
theory. 
\end{quote}
For a realist interpretation:
\begin{quote}
It {\em is} appropriate in quantum theory to say that an individual system
possesses values of its physical quantities. In this context, 'appropriate'
signifies that propositions of this type can be handled using standard
propositional logic. 
\end{quote}

To understand this text, one has first to know what are 'physical quantities
of an individual quantum system'. For practically all physicists, these are
the values of observables (examples are
\cite{Isham,d'Espagnat,Peres,Bub}). For a realists, it remains then only to
ask what can be a maximal set of {\em determinate} (but perhaps unknown)
properties for a system, $S$, say, in a state $\rho$, the so-called
determinate set. The assumption that they are determinate must not lead to any
contradictions that might follow from the standard quantum mechanics and
possible measurements of other observables in the determinate set as well as
from some further desirable conditions (for details, see \cite{Bub}). Examples
of such contradiction can be obtained from the Kocher-Specker theorem (see,
e.g., \cite{Peres}). Important is that the absence of contradiction is only
required for measurements of observables inside the determinate set and
measurements of other observables can be ignored. The resulting mathematical
problem has been solved (see \cite{Bub}). All sets of determinate properties
are necessarily restricted in the sense that they never contain all
observables that are in principle measurable on $S$ in $\rho$ (at least for
Hilbert spaces that have more than two dimensions). 

The standpoint of the present paper is very different. We consider a value of
observable $O$ that can be measured on a quantum system $S$ as a physical
quantity of composite system $S+A$, where $A$ is the apparatus that measures
$O$. For us, it is, therefore, not a physical quantity of individual system
$S$ and we call it {\em extrinsic property} of $S$. Probabilities can be
interpreted epistemically only as concerning the composite system. The
so-called determinate properties are then mostly the extrinsic ones. Extrinsic
properties of $S$ have an important but only intermediate role: from them,
something about genuine physical quantities of $S$ can be inferred. We call
the part of standard quantum mechanics that concerns extrinsic properties {\em
  phenomenology of observation}. A precise account is Sec.\ 2. 

The main new idea of the present paper is a proposal of what the quantities of
individual quantum systems are. It is very different from what is taught at
universities and what is believed by experts who know all existing
literature. We start from two principles. First, our reality condition is
stronger than that for determinate properties: it requires that the
attribution does not lead to contradictions with {\em any} possible
measurement that could be performed on $S$ in $\rho$ according to the basic
rules of standard quantum mechanics. Hence, most determinate properties
violate our reality condition. Second, we necessarily consider properties that
are different from values of observables. They are more sophisticated and need
not have their values in $\mathbb R^n$. If we find such a quantity, we accept
the hypothesis that does ascribe it to individual quantum systems and call
such properties {\em intrinsic}\footnote{We take the standpoint that such
  ontological hypotheses are meaningful if they have some relation to
  observation. An ontological hypothesis is allowed if its consequences are
  not disproved by existing evidence and if it is logically compatible with
  other physical theories.}. In Sec.\ 3, it is shown that there are two kinds
of intrinsic properties, structural and conditional, and that there is plenty
of them. 

Many physicists are aware of objectivity of properties such as mass and charge
but our notion of {\em structural properties} includes much more quantities. A
structural property of any quantum system is, e.g., the form of its
Hamiltonian operator. Clearly, it is a specific feature different from system
to system and it is amenable to an exact mathematical description even if not
by a quantity that takes on numerical values. According to our opinion, the
ultimate aim of all quantum measurements (such as the scattering experiments
in CERN) is to determine the structural properties of real quantum systems
(such as parameters of the standard model). {\em Conditional} are those
properties that can be given to quantum systems by preparations. For example,
Dirac (\cite{Dirac}, P. 46) and von Neumann (\cite{JvN}, P. 253) added to the
standard quantum mechanics the interpretation of eigenstates as the only
special case in which a value of an observable is determinate before
measurement. But our notion of conditional properties goes far beyond
this\footnote{The analysis of experiments in the so-called {\em quantum
    mechanics on phase space} \cite{Schroeck} introduces the notion of
  property that generalizes single values of observables to probability
  distributions of such values. Some of these properties are similar to some
  of our conditional properties. However, quantum mechanics on phase space is
  a modification rather than an interpretation of quantum mechanics. It
  postulates the existence of informationally complete measurements. Our
  approach is different.}. Thus, the basic ones are summarized and
mathematically expressed by state operators, while the more advanced ones
extend the information given by state operators. Existence of such advanced
properties has been established by our previous paper \cite{TH}. 

Our paper is the first systematic, self-consistent and complete account of
structural and conditional properties. It makes the ontological hypotheses of
objective existence of these properties and recognizes that the hypotheses
form an interpretation of quantum mechanics which is different from any other
known one (versions of Copenhagen, Everett, etc.). The interpretation may be
called realist because it satisfies Isham's criterion. 

As mentioned above, standard quantum theory works directly only with classical
systems and classical properties. In particular, the measuring apparatuses are
classical systems. In this context, another classification criterion of
anti-realist versus realist interpretation emerges. Measurement is considered
as a fundamental notion of the theory and classical systems as different from
quantum systems in the former and classical systems are considered as quantum
systems with special quantum properties and measurement processes as some kind
of quantum processes that represent nothing fundamentally new in the latter
(see, e.g., Isham, P. 68). 

Sec.\ 4 turns to this part of our interpretation. We conjecture that the
dichotomy of quantum and classical worlds can be replaced by the difference
between extrinsic and intrinsic properties of quantum systems. In particular,
average values of observables in prepared states are intrinsic. This enables
us to use methods of statistical physics to construct quantum models of some
classical properties. An example explaining this idea is carefully described
in Appendix A. A tentative generalisation to the classical properties of a
macroscopic body that determine its complete classical state is given: they
can be chosen as intrinsic average values and small variances of certain
operators of the underlying quantum system. Recently, there has been some
progress concerning quantum models of further classical properties
\cite{hajicek}. 

Our interpretation also gives some non-trivial hints of how the measurement
problem is to be approached. On the one hand, corresponding to its notion of
classical systems and classical properties, the assumption that measuring
apparatuses are exclusively classical (that is non quantum) has to be
abandoned. On the other, von Neumann model of measuring apparatus, where
readings of an apparatus are eigenvalues of an operator, is incompatible with
it.

\section{Phenomenology of observation}
This section briefly describes what part of standard quantum theory belongs to
the phenomenology of observation. To shorten the exposition, we shall base it
on the accounts of standard quantum mechanics as given in
\cite{Isham,d'Espagnat}. The Rules will be referred to just by their numbers. 

At the beginning of any measurement stands what is usually called {\em
  preparation}. The name is somewhat misleading. What is meant is a set of
conditions that can be described in classical terms, to which an individual
quantum system has been subject and that determine its state. This can, but
need not, include some human activity in laboratory. For example, we can know
that a quantum system inside the Sun is the plasma with a given composition
and that its classical conditions are certain temperature and pressure. The
description of classical conditions is important in order that the same
preparation can be recognized or reproduced. Thus, a series of repeated
experiments is feasible, and the set of individual quantum systems obtained by
repeating the same preparation is called {\em ensemble}. Clearly, the notion
of ensemble is in many aspects closely connected to that of
preparation. Isham's Rules 1 and 2 and d'Espagnat's Rules 1, 2 concern
states. 

At the end of any quantum measurement there is what is often called a {\em
  registration}. It is an interaction of an individual quantum system in a
specific state with a classical system, the {\em measuring apparatus} that
determines a value of a set of commuting observables. About representation of
observables by operators is the Rule 3 by Isham and Rules 4, 5 and 6 by
d'Espagnat\footnote{A more general mathematical object can be mentioned as
  representing registrations, the {\em positive operator valued measure}
  (POVM). However, any registration represented by a POVM of a system $S$ is
  nothing but a registration associated with a suitable observable of an
  extended system, $S+S'$, $S'$ being the so-called ancilla
  \cite{Peres}. Thus, conceptually, POVM belong to extrinsic properties
  because of both measuring apparatus and ancilla.}. 

Rule 4 by Isham and Rule 3 by d'Espagnat concerning dynamics are considered as
a part of standard quantum mechanics but not of the phenomenology. All
mathematics that is associated with the included Rules, such as the theory of
Hilbert spaces and self-adjoint operators, are considered as part of the
phenomenology. 

Let us introduce the word 'property' in order to have a general notion of
observable characteristics concerning quantum systems. For instance, the
values observables in quantum mechanics are properties. We define:   
\begin{df}
Extrinsic properties of quantum system $S$ are those values of
observables\footnote{More generally, extrinsic properties can be described as
  linear subspaces in the Hilbert space of the system. They represent the
  mathematical counterpart of the so-called YES-NO experiments
  \cite{Piron}. The set of linear subspaces admits the usual operations on
  conjunction (linear hull), disjunction (intersection) and negation
  (orthogonal complement), but the resulting orthocomplemented lattice is not
  a Boolean lattice \cite{BvN}. As it is well known, the set of 'classical'
  properties of a single system forms a Boolean lattice (of subsets of
  classical phase space). If we pretend that the extrinsic properties of a
  quantum system are properties of a well-defined single system, then we are
  lead to abandon the ordinary logic and introduce the so-called {\em quantum
    logic}. But this pretence is against all logic because the extrinsic
  properties are properties of many different systems each consisting of the
  quantum system plus some apparatus.} pertaining to $S$ that are not uniquely
determined by the preparation of $S$. 
\end{df}

The extrinsic properties are not real properties of quantum systems in the
following sense: the assumption that an extrinsic property $P$ of a quantum
system $S$ as measured by an apparatus $A$ is possessed by $S$ independently
of, or already before, their registration, leads to contradictions with other
possible measurements on $S$. An example is the well-known double-slit
experiment (see also \cite{d'Espagnat},Sec.\ 4.3).

\section{The intrinsic properties of quantum systems}
A property $P$ can be ascribed directly to a quantum system $S$ if
consequences of $S$ possessing $P$ do not contradict results of any
measurement that can be carried out on $S$ (even very difficult measurements
so as to be practically not feasible). Let us define: 
\begin{df}
Let $S$ be a quantum system and $P$ a property that can be directly ascribed
to $S$ alone so that the assumption of $S$ really possessing $P$ does not lead
to contradiction with any measurements that can in principle be done on $S$
according to the rules of standard quantum mechanics. Then $P$ is called
intrinsic property. 
\end{df}
For understanding the notion, it is important to discriminate between
attributing a property to {\em an individual quantum system} on the one hand
and directly measuring the property on {\em an individual quantum system} on
the other. Many structural properties, such as cross sections or branching
ratios, are obtained only after many measurements, of rather different
(extrinsic) properties, on many copies of a system. The same is true for
conditional properties such as a state operator. Still, no formal-logical
problem arises if one wants to attribute them to individual systems.

\subsection{Structural properties}
First, we turn to those intrinsic properties that are easy: nobody would
seriously deny that they can be ascribed to quantum systems. They are also the
most important properties of quantum systems in the sense that the ultimate
aim of experimenters is to determine them. 

Quantum systems can be classified into equivalence classes with the beautiful
property that the structures of each two different systems of the same class
are absolutely identical. Examples: electrons, protons, hydrogen atoms,
etc. Let us define: 
\begin{df}
All properties that are uniquely determined by the class of quantum system are
called  structural. 
\end{df}

The first among the structural properties is the {\em composition of a quantum
  system}. Experience and practice lead to ideas about what such a composition
can be. For example, in the non-relativistic case, there must be a definite
number\footnote{There are non-relativistic systems, in which some particle
  numbers are variable, such as those of quasi-particles in solid state
  physics. Of course, these particle numbers do not belong to the structure of
  the systems and they are not intrinsic but extrinsic properties in our
  conception.} of some particles with definite masses, spins and charges. For
a relativistic case, there are analogous rules: we have fields of certain
(bare) masses, spins and charges. For example, the non-relativistic model of
hydrogen atom consists of two particles, proton and electron, that have
certain masses, spins and charges. 

The next step is to determine the quantum observables that can be measured on
the system. For example, each particle contributes to the observables by three
coordinates and three momenta. Thus, in the hydrogen case, there will be (in
addition to other observables) six coordinates and six components of
momenta. The {\em set of observables that can be measured on a given system}
is its intrinsic structural property. This information is different from that
about the values of these observables\footnote{More precisely, the set of
  observables can be embedded in a so-called $C^*$-algebra that represents a
  part of the physical structure of the system, see \cite{thirring}, Vol.\
  3. Thus, it is an intrinsic property of it. Moreover, such algebras have
  representations on a Hilbert spaces. A representation defines the Hilbert
  space of the system. Of course, for systems with finite number of degrees of
  freedom, the Hilbert space representation is uniquely defined (up to unitary
  equivalence) by the algebra, so it does not contain any further information
  on an independent structure of the system, but the algebras of relativistic
  fields possess many inequivalent representations of which only few are
  physical, corresponding to different phases of the system. A physical
  representation is then clearly an independent structural intrinsic property
  of the field.}. The algebras of observables contain for example information
about superselection observables (which form the centre of the algebra), so
these observables are also structural properties. 

The composition and the observables of a system are used to set up the
Hamiltonian of non-relativistic or the action functional for relativistic
systems. The {\em form of the Hamiltonian or the action} are mathematical
expressions of the structure and thus intrinsic properties\footnote{The energy
  of a system $S$ that can be measured by suitable apparatus $A$ is an
  observable. The value of energy obtained on $S$ by $A$ is a 'beable', it is
  not an intrinsic property of $S$ but that of the composite system $S +
  A$. The three notions of measured energy value, energy measurements and the
  form of Hamiltonian are related to each other but they are clearly not
  identical.}. 

Using the Hamiltonian or the action, we can write down the dynamical laws: the
Schr\"{o}dinger equation or the path-integral formula. Hence, the dynamical
part of standard quantum theory (Rule 4 by Isham, Rule 3 by d'Espagnat) is
included into the structural properties of our interpretation. From the
dynamical laws, other important intrinsic properties can be calculated, for
example the spectrum of the hydrogen atom. The spectrum is clearly a
structural intrinsic property of the hydrogen atom that can be ascribed to the
system itself independently of any measurement. This will not lead to any
contradictions with other measurements or ideas of quantum mechanics. We can
recognize the system with the help of its intrinsic properties. For example,
if we detect light from somewhere deep in the Universe and find the Balmer
series in its spectrum, then we know that there is hydrogen there. The numbers
such as cross sections, branching ratios etc. are further examples of
structural intrinsic properties. Moreover, the Hamiltonian contains
information about all symmetries of the system, because these are represented
by the operators that commute with the Hamiltonian. Thus, symmetries are
structural properties. 

Next, it seems that many-particle systems may have structural properties that
are not found in small quantum systems. An example is provided by molecules of
the deoxyribonucleic acid. The number of their structures grows (roughly)
exponentially with the number of the four kinds of constituents because
possible orderings of the constituents define different structures. It is
clearly wrong to say that we know all kinds of structural properties of
macroscopic systems and investigations in this direction might be useful. For
example, rich intrinsic properties of large systems might imply the observed
properties of the universe and so enable a new approach to quantum cosmology
without need of bizarre theories such as many world interpretation
\cite{Isham}. 

These steps form the everyday practice of quantum mechanics. An application of
quantum mechanics  starts with a proposal of a model for the quantum system
under study. This is done by specifying its structural properties. For each
system, we can attempt different possible models, calculate the extrinsic
properties of each and compare with the experimental evidence gained in a
number of quantum measurements. In this way, the models can be confirmed or
disproved. The sets of intrinsic and extrinsic properties are model
dependent. What is relevant is that every quantum model exhibits both
intrinsic and extrinsic properties.

\subsection{Conditional properties}
Encouraged by the triviality of the assumption that structural properties are
intrinsic, we start to look for some intrinsic properties that can have
different values for one and the same class of quantum systems. Let us define: 
\begin{df}
A property is conditional if its value is uniquely determined by a preparation
according to the rules of the standard quantum mechanics. The 'value' is the
value of the mathematical expression that describes the property and it may be
more general than just a real number. No registration is necessary to
establish such a property but a correct registration cannot disprove its
value; in some cases, registrations can confirm the value. 
\end{df}

This can best be explained by examples. Suppose that a system $S$ has been
prepared in the eigenstate $|o\rangle$ of an observable $O$ with the
eigenvalue $o$. Now, think: could any conceivable registration made on $S$
thus prepared in $|o\rangle$ contradict the assumption that $S$ possesses the
value $o$ of $O$? The standard rules of quantum mechanics clearly say no. More
generally, any quantum state $\rho$ that has been prepared for the system $S$
is a property of $S$; its value $\rho$ (i.e., a positive self-adjoint operator
with trace 1) can only be confirmed by registrations following the
preparation. 

Next consider a particle $S$ with spin 1/2. The state with spin projection to
the $z$-axis equal to $\hbar/2$ can be prepared. Then, no contradictions can
result from the assumption that $S$ with this value of $\sigma_z$ really
exists. Thus, the value $\hbar/2$ of $\sigma_z$ is one example of a
conditional property. Of course, $S$ does not possess any value of $\sigma_x$
after such preparation; this would only be brought about by a corresponding
measurement and is not uniquely determined by the preparation. Hence, it is an
extrinsic property. On the other hand, the average (also called expectation or
mean) value of $\sigma_x$ in the prepared state has a well-known value defined
by the state and hence it is another example of a conditional
property. (Averages will play a key role in the definition of classical
properties.) 

One could try to object that there has been the preparation, this is a 'kind
of measurement' and the property depends on this 'measurement'. Moreover, the
preparation has used an apparatus $A$, say, and the property seems therefore
to be a property of the system $S+A$ and not $S$ alone. However, these
objections concern clearly also the structural properties: an apparatus that
prepares a beam of electrons is different from that for a beam of
protons. Moreover, they could be also raised in Newton mechanics: giving a
snooker ball momentum $p$ requires a careful action of the cue. Still, nobody
questions the existence of the momentum $p$ on the ball alone after the poke. 

Any preparation defines a specific state of the quantum system. States can be
pure ones or mixtures and we describe them generally by state operators.
There is a measure of how restrictive and special the preparation process is,
namely the entropy\footnote{The term 'entropy' always means the von Neumann
  entropy in this paper.}. It is a function of state and thus an important
conditional property. 

More advanced examples of conditional properties concern mixtures. In
\cite{TH}, we have shown that, in some cases, $\rho$ does not contain all
information available by registrations concerning the prepared ensemble. One
can consider the ensemble of particles as defining a state of a single
particle, of two particles, etc., and the information involved in the
preparation can thus be described by a set of state operators. Such a set is
the mathematical description of the property, which is clearly intrinsic, and
of conditional character. It can be confirmed by registration. 

An example of conditional property that cannot always be confirmed by
registration is the difference between proper and improper mixtures (for
definitions, see \cite{d'Espagnat}). Suppose that a physicist prepares states
$|1\rangle,\cdots,|n\rangle$ of a quantum system $S$ and mixes them with
frequencies $c_1,\cdots c_n$ so that the resulting state of $S$ can be
described by state operator 
\begin{equation}\label{properm}
  \rho = \sum_{k=1}^n c_k|k\rangle\langle k|.
\end{equation}
This is a proper mixture and the particular decomposition (\ref{properm}) of
$\rho$ is a conditional property, that is a real property of the prepared
ensemble. In particular, we can assume that the system {\em really is} always
in {\em either} of the states $|1\rangle,\cdots,|n\rangle$ with respective
probabilities $c_1,\cdots, c_n$. The decomposition (\ref{properm}) is not the
unique decomposition that the state operator $\rho$ admits but it is the one
that is uniquely determined by the preparation. According to the standard
quantum mechanics, which we adhere to, two different decompositions of the
same state operator cannot be distinguished by any registrations. This however
represents no embarrassment for us: we do not adhere to the positivist maxim
that there is only what is measured. It is also clear that any time evolution
of such a proper mixture is the proper mixture of the evolved states with the
same probabilities and determines a well-defined decomposition at each
time. On the other hand, given an improper-mixture state $\sigma$ of $S$ then
no individual system $S$ can in general be assumed to {\em really be} in any
of the component states of any decomposition of $\sigma$. An example is the
Einstein-Podolsky-Rosen experiment \cite{Peres}. 

What are possible conditional properties of a given quantum system? Clearly,
all state operators that can be prepared (some limitations are due to
superselection rules) belong to them. Other properties are either derived from
the state operators (such as the average of an observable) or added to state
operators (such as the decomposition describing a proper mixture). It seems,
that state operators are also universal in the following sense. Even if we do
not know what the source or origin of a system is (for example the protons
coming in cosmic radiation), the assumption that it is described by some state
operator does not lead to any contradictions. 

All examples that have been listed show that the intrinsic and extrinsic
properties are physically inseparably entangled with each other. Even the
definition of intrinsic properties uses the notion of registration of
extrinsic properties: an intrinsic property can be ascribed to the system
alone without giving rise to contradictions with the results of all possible
registrations. Similarly, extrinsic properties cannot be defined without the
notion of a measuring apparatus with its classical properties, which are a
kind of intrinsic properties in our point of view (see the next
section). Thus, e.g., the notion is clearly untenable that the intrinsic
properties can be explained purely in terms of the extrinsic ones. Still, both
kinds of properties are logically clearly distinguished, and we conjecture
that the physical in-and-extrinsic tangle does not lead to any logical
contradictions.

\section{Classical properties}
There is a lot of systems around us that behave as classical physics
prescribes, at least to a good approximation. We would like to have quantum
models of such systems. The features that are most difficult to reproduce are
summarized in the so-called principle of {\em macroscopic realism}
\cite{Leggett}, but as formulated by Leggett it is too strong for our
needs. Let us modify the principle as follows: 
\begin{enumerate}
\item A macroscopic system which has available to it two or more {\em distinct
    classical states} is at any given time in a definite one of those states. 
\item It is possible in principle to determine which of these states the
  system is in without any effect on the state itself or on the subsequent
  system dynamics. 
\end{enumerate}
Here, we have just replaced Leggett's 'macroscopically distinct (quantum)
states' by 'distinct classical states,' and we call the resulting principle
{\em modified macroscopic realism}. Of course, if the classical states include
pure quantum states, point 1 of the macroscopic realism violates the principle
of superposition. Then, one has to assume that some as yet unknown phenomena
exist at the macroscopic level which are not compatible with standard quantum
mechanics (see, e.g., \cite{Leggett} and the references therein). However, no
such phenomena have been observed. We ought therefore to suggest how our
modified macroscopic realism could be derived from quantum mechanics, at least
in principle. 

Observe that the modified macroscopic realism as it stands cannot be obtained
from the decoherence theory, at least in its present form. For that e.g.\ the
word 'is' in point 1 had to be replaced by 'appears to be' (see
\cite{zeh}). The 'appears' would undermine our quantum realism. We are
optimistic because it seems that our interpretation provides some new
tools. Of course, this does not mean that the argument is circular but only
that the output does not contradict the input. 

Hence, let us assume that all physical systems are quantum systems. More
precisely, there is one level of description (approximative model of some
aspects of a real system) of a classical system $S_c$ and of its classical
properties for which quantum theory is not needed, namely the classical
description, and for which the modified principle of macroscopic realism is
valid. In addition, every classical system $S_c$ can also be understood as a
quantum system $S_q$ underlying $S_c$ such that the classical properties of
$S_c$ are some intrinsic properties of $S_q$. This follows from our definition
of intrinsic properties and from the modified macroscopic realism. 
Namely, any classical state of $S_c$ is defined by values of some classical
properties. As it must also be a state of $S_q$, the assumed reality of the
state requires that these classical properties are intrinsic properties of
$S_q$. 

The quantum description of $S_q$ consists of the following points. 1) The
composition of $S_q$ must be defined. 2) The algebra of observables that can
be measured on $S_q$ is to be determined. As any observable is measurable only
by a classical apparatus, the existence of such apparatuses must also be
assumed from the very beginning. Quantum description of $S_q$ will thus always
contain some classical elements. This does not mean that classicality has been
smuggled in because, in our approach, classical properties are specific
quantum ones. 3) A Hamiltonian operator or an action functional of the system
must be set up. Finally, the known classical properties $P_1, P_2, ..., P_K$
of $S_c$ must be listed and each derived as an intrinsic property of $S_q$
from the three sets of assumptions above. This is a self-consistent framework
for a non-trivial problem. 

There are intrinsic properties of $S_q$ that are not classical properties of
$S_c$, e.g., the set of all quantum observables measurable on $S_q$. Hence,
classical properties must be some specific intrinsic properties and the
question is, which. 

To begin with, let us consider the so-called semi-classical (or WKB)
approximation. This includes the observation that, for a number of systems,
the average values in special states of a number of quantum observables follow
classical (say, Newton mechanics) trajectories. This is surely a good start
because, as we have seen in Sec.\ 3.2, in some cases, average values can be
considered as conditional properties. Moreover, everything what we can measure
on classical systems has a form of average value and its variance. This is
evident from the description of any classical experiment. How are these
averages and variances related first to the relevant classical theory and,
second, to the averages and variances of quantum operators? 

As the first question is concerned, it is often assumed that improvements in
measuring techniques will in principle, in some limit, lead to zero
variance. This is in agreement with the classical theory such as Newton
mechanics. It predicts that the trajectories are completely sharp if the
initial data are so, and does not put any limit on the accuracy with which the
initial state can be prepared. The point of view adopted here is different (it
is originally due to Exner \cite{Exner}, p.~669, and Born \cite{Born}): some
part of the variances can never be removed and the classical theories are only
approximative models. 

The second question contains two different problems. First, if we measure
several times the position of the Moon on its trajectory around the Earth,
then the variance in the results is surely not connected to our preparing the
system of Earth and Moon these many times to get the desired ensemble. But the
classical systems are robust in the sense that most classical measurements
practically do not disturb them (point 2 of macroscopic realism). Thus, one
can assume that the values we obtain by several measurements on one and the
same system could equivalently be obtained if the measurements were performed
on several identically prepared systems. The hypothesis is therefore plausible
that some intrinsic properties we are looking for are averages with small
variances associated with preparations under identical relevant conditions. If
the variance of a given average value is sufficiently small, it can be and is
usually viewed as a property of each individual element of the ensemble. We
conjecture that this is the way classical systems come to possessing their
properties just from the classical experimental point of view. Second, there
are classical properties that cannot be viewed as intrinsic averages of
quantum operators but are structural or different conditional properties. Some
examples will be given later. 

Next, there is a restriction on quantum models of classical properties: they
cannot be averages with small variance that are defined by pure states such as
coherent ones. Not only are pure states readily linearly superposed but any
quantum registration (a generalized measurement: positive operator valued
measure) that were to find the parameters of a coherent state would strongly
change the state. The only available hint of what classical properties may be
comes from thermodynamics. Indeed, statistical physics is a successful method
of deriving macroscopic properties from microscopic ones. Moreover, the
notions of structural and conditional properties enable a cleaner formulation
of quantum statistical physics. The following is a brief sketch of a specific
example from the thermodynamic-equilibrium theory. 

Let $S$ be a (non-relativistic) quantum system with number of particles
comparable to Avogadro number. We call such systems {\em macroscopic}. Let its
structure be described by a Hamiltonian $H$. Imagine that $S$ is prepared in
all possible quantum states (not necessarily by humans  in
laboratories). Consider only those of these states that have a fixed average
value $\bar{E}$ of internal energy. A well-defined average value is a
conditional property that exists for each of the prepared states and hence the
imaginary selection (without need of any additional registration) is
legitimate. Let us call this subset of prepared states $\bar{E}$-{\em
  ensemble}. 

Next, let the state $\rho_{\bar{E}}$ be defined by the requirement that it
maximizes the entropy under the condition that the average internal energy has
the value $\bar{E}$. This is known as the Gibbs state of $S$. The state is
purely mathematical because no preparation process for it has been specified.
The central conjecture of statistical physics reads: {\it For macroscopic
  systems, important statistical properties of $\bar{E}$-ensemble coincide to
  a very good approximation with the corresponding statistical properties of
  $\rho_{\bar{E}}$.}  Claims, equivalent to this conjecture can to a large
extent be derived from quantum mechanics (\cite{thirring}, Vol.\ 4), in the
thermodynamic limit. Bayesian approach \cite{Jaynes} to probability and
entropy is also helpful. (The thermodynamic limit is, of course, not a
physical condition but a mathematical method of how the structural property of
being macroscopic can be brought into play.)

What are the 'important' statistical properties above? Some of them are
average values and variances of a very small but definite subset ${\mathcal
  T}_S$ of the algebra of all quantum observables of $S$. Clearly, these are
conditional properties because they are determined by the prepared states from
$\bar{E}$-ensemble. The observables from ${\mathcal T}_S$ are extensive
quantities associated with some of the ordinary thermodynamic variables. For
instance, consider a gas in a vessel of volume $V$. The operator of internal
energy $\mathbf E$  of the gas (the Hamiltonian in the rest frame) belongs to
${\mathcal T}_S$. We can also choose some small but macroscopic partial volume
$\delta V$ at a specific position within $V$ and consider the particle number
$\delta\mathbf N$ inside $\delta V$. Operator $\delta\mathbf N$ can be
constructed from the projectors on the position eigenstates of all particles
in $S$. The energy $\delta\mathbf E$ inside $\delta V$ can be constructed as a
coarse-grained operator (see \cite{kampen}) because the exact energy operator
does not commute with operators of particle positions. It seems that all
quantum observables from ${\mathcal T}_S$ are macroscopic in the sense that
they have a coarse-grained character or concern many particles. 

Other thermodynamic variables are not average values of quantum
observables. Examples are structural quantities such as the total mass and
particle number of a macroscopic body or state quantities such as the maximal
value of entropy and the corresponding temperature (the Lagrange multiplier
that appears naturally in the problem of maximization of entropy). 

The average values of observables from ${\mathcal T}_S$ determine a
thermodynamic state of the system. Let us consider such a state as an example
of a classical state (appearing in point 1 of the Principle) of the quantum
system $S$. For example, the internal energy and the volume determine the
state of a simple ideal gas. Thus, one macroscopic state is compatible with a
huge number of microscopic (quantum) states of $S$. It is very important to
understand that a macroscopic state of $S$ is conceptually different from any
microscopic state of it, and that there are no linear superpositions of
macroscopic states. The sets of average values of operators from ${\mathcal
  T}_S$ do not form a linear space that could lead to a definition of state
superposition of a fixed system: addition of extensive quantities entails
addition of the corresponding systems. 

It can be shown that the observables from ${\mathcal T}_S$ have negligible
relative variances in the Gibbs state. (The property that they are extensive
plays an important role.) Thus, the average values of the observables can be
given individual meaning: each individual system from $\bar{E}$-ensemble {\em
  possesses} a value of the observables within certain limits. Is such an
average already a classical property satisfying the requirements of the
macroscopic realism? Point 1. is satisfied by construction. Point 2. is just
plausible as yet: the influence of measurement can still be large as concerns
the microstate but it can change it to another microstate that is compatible
with the original macrostate and so it need not change the
macrostate. Clearly, statistical physics in our interpretation is the quantum
theory of at least some macroscopic properties. 

The discussion above motivates a general definition of the classical state
making it analogous to the thermodynamic state as follows.  
\begin{df}
Let the state of classical system $S_c$ be described by the set of $n$ numbers
$\{A_1,\cdots,A_n\}$ that represent values of some classical observables with
variances $\{\Delta A_1,\cdots,\Delta A_n\}$. Let the corresponding quantum
system $S_q$ contain in its algebra a set of $n$ observables $\{{\mathbf
  a}_1,\cdots,{\mathbf a}_n\}$ that correspond to the classical ones. Then the
quantum counterpart of the classical state $\{A_1,\cdots,A_n\}$ is the set of
all quantum states $\rho$ such that 
$$
\text{Tr}(\rho {\mathbf a}_k) = A_k, \quad \sqrt{\text{Tr}(\rho {\mathbf
    a}_k^2)-[\text{Tr}(\rho {\mathbf a}_k)]^2} = \Delta A_k. 
$$
\end{df}
Let us call the quantities in ${A_1,...,A_n}$ state coordinates\footnote{They
  are not uniquely determined by the classical system and we assume that the
  choice can be done so that all state coordinates are averages of quantum
  operators.}. For example, the classical state of a mass point in mechanics
can be described by three coordinates $Q_k$ and three momenta $P_k$, and the
operator algebra of its quantum analogue contains the corresponding operators
${\mathbf q}_k$ and ${\mathbf p}_k$. Of course, we can define a sensible
classical state only for macroscopic systems so that their classical states
contain huge numbers of quantum states and in this way much less information
than their quantum states. Each of the many quantum states satisfying the
above equations can be viewed as representing one and the same classical
state. 

Classical states defined in this way can be understood as equivalence classes:
two quantum states are equivalent, if the state coordinates have the same
averages and variances in them. For such classes, one can try to define a
superposition operation by forming superpositions of vectorial representative
of the states: let $|a\rangle \in \{\rho\}$ and $|b\rangle \in \{\sigma\}$,
where $\{\rho\}$ denotes the class with representative $\rho$, then 
$$
\{c|a\rangle + d|b\rangle\} := c\{\rho\} + d\{\sigma\}\ .
$$
However, we find that there are often more vectorial representations in each
class and that superposition of another pair does not lead to the same
class. $|a'\rangle \in \{\rho\}$ and $|b'\rangle \in \{\sigma\}$ are other
such vectors then  
$$
\{c|a\rangle + d|b\rangle\} = \{c|a'\rangle + d|b'\rangle\}
$$
can hold only in few exceptional cases.

To summarize the main points of our theory of classical properties, let us
first compare it with some well-known approaches to the problem. Thus, we
mention the quantum decoherence theory \cite{zurek, zeh}, the theories based
on coarse-grained operators \cite{Peres,poulin,Kofler}, the Coleman-Hepp
theory \cite{Hepp,Bell3,Bona} and its modifications \cite{Sewell}. At the
present time, the problem does not seem to be solved in a satisfactory way,
the shortcoming of the above theories being well known \cite{d'Espagnat,
  bell2,Wallace}. Our approach is free of these shortcomings. 

It starts at the idea that all classical properties of a macroscopic system
$S$ in a quantum state $\rho$ are certain {\em intrinsic} quantum properties
of $S$ in $\rho$. Then, first, intrinsic properties are quantum properties of
all quantum systems and there is no question about how they emerge in quantum
mechanics. This avoids e.g. the artificial construction in the Coleman-Hepp
approach. The new point is that they are considered as, and proved to be,
objective in our paper. Hence, second, they could in principle serve as
classical properties because they can satisfy the principle of classical
realism. This avoids the problems of both the quantum-decoherence and the
coarse-grained theory that assume values of quantum observables to be real.
This, as analysed in \cite{Bub}, can be done only for restricted classes of
observables, all other measurements being forbidden. Third, we conjecture that
certain macroscopic quantum systems possess intrinsic properties that can
model all their classical properties. Hence, classical states and properties
defined in the present paper are available only for some quantum systems and
the relation between classical and quantum states is not one-to-one but
one-to-many. This is different from other approaches such as Wigner-Weyl-Moyal
scheme, quantum-mechanics-on phase-space theory or coherent state approach.
Finally, our modelling or construction of classical properties uses the way
analogous to that of statistical physics. Thus, definition 5 starts a project
of modelling classical properties of quantum systems including the internal
(thermodynamic) and external (mechanical) properties. A full derivation
including the complete list of all assumptions is described with the help of
an example in Appendix A. Models of classical mechanics are constructed in
\cite{hajicek}.

An important piece of our interpretation is the existence of classical
macroscopic apparatuses that are needed for the phenomenology of
observation. Some necessary conditions such apparatuses must satisfy not only
in order that the phenomenology works but also that our realist interpretation
has a reliable basis are summarised in the modified principle of macroscopic
realism. 

The problem to construct a quantum model of registration process is the most
difficult one in the field of conceptual foundation. A quantum explanation of
classical properties is only a part of the problem. There is much activity in
this field. The references given above deal also with the measurement
problem. No satisfactory solution seems to be known. 

One cause of the difficulties may be the model of measuring apparatus that has
been proposed by von Neumann and by Jauch \cite{JvN,Jauch}. The key assumption
of the model is that the values shown by the apparatus are some of its
extrinsic properties. For example, the pointer states are eigenstates of some
quantum operator. Let us briefly describe it for the case of quantities with
discrete values (continuous quantities would need a slightly different
approach). 

Suppose quantum system $\mathcal S$ is prepared in initial state
$|{\mathcal S}1\rangle$ and the observable to be measured, $\mathbf a$, has
eigenvalues $a_k$ and eigenstates $|a_k\rangle$. We can write 
$$
|{\mathcal S}1\rangle = \sum_k c_k|a_k\rangle.
$$
The apparatus that makes the registration is quantum system $\mathcal A$ in
initial state $|{\mathcal A}1\rangle$ and its pointer observable $\mathbf A$
has eigenvalues $A_k$ and eigenstates $|A_k\rangle$. 

The next assumption is that there is an interaction between the two systems
that leads to unitary evolution 
$$
|{\mathcal S}1\rangle\otimes|{\mathcal A}1\rangle \mapsto \sum_k
c_k|a_k\rangle\otimes|A_k\rangle. 
$$
If we trace out $\mathcal S$, we obtain the final state of the apparatus,
\begin{equation}\label{Af}
\sum_k |c_k|^2|A_k\rangle\langle A_k|.
\end{equation}
This a mixture of the eigenstates $|A_k\rangle\langle A_k|$ of the apparatus
with the 'correct' probabilities $|c_k|^2$.  

The problem is that (\ref{Af}) is not a proper mixture. Nothing prevents us to
view the process above as a preparation of the apparatus in the state
(\ref{Af}) and this preparation does not contain any steps that would bring
$\mathcal A$ into any of the states $|A_k\rangle$ during each individual
measurement\footnote{In the decoherence theory, another component, the
  environment, is added at the beginning  and traced out at the end. The
  result is again an improper mixture and the problem remains exactly the
  same.}. The usual way out is to employ another apparatus $\mathcal B$, which
is non-quantum so that an interaction between $\mathcal A$ and $\mathcal B$
can bring $\mathcal A$ always into one of the states $|A_k\rangle$. 

Clearly, the model contradicts the experience. One apparatus is sufficient, it
is a system with classical properties and the outcome of any individual
measurement is represented by a definite change of a classical property of the
apparatus. Moreover, the model is incompatible with our interpretation. The
extrinsic property that represents the apparatus readings had to be replaced
by an intrinsic one. We have to use our theory of classical properties as
described in the previous section. To construct a model of such an apparatus
is a problem that will be addressed in a separate paper. 

To summarize: Our interpretation suggests a new approach to quantum theory of
classical properties and of measurement because it allows quantum systems to
have also properties that are not extrinsic.

\appendix
\section{Quantum model of classical property}
The purpose of Appendix A is to construct a quantum model of a classical
property, the length of a body, as an average value with a small variance. No
original calculation is to be expected, but simple and well known ideas are
carefully interpreted according to the lines described in Sec.\ 4. This
entails that, first, the quantum structure of the system must be defined,
second, the basic intrinsic properties such as the spectrum calculated, and,
third, some intrinsic properties derived that satisfy our definition of
classical property.

\subsection{Composition, Hamiltonian and spectrum}
We shall consider a linear chain of $N$ identical particles of mass $\mu$
distributed along the $x$-axis with the Hamiltonian 
\begin{equation}\label{Ham}
 H = \frac{1}{2\mu}\sum_{n=1}^N p_n^2 + 
\frac{\kappa^2}{2}\sum_{n=2}^N (x_n - x_{n-1} - \xi)^2, 
\end{equation}
involving only nearest neighbour elastic forces. Here $x_n$ is the position,
$p_n$ the momentum of the $n$-th particle, $\kappa$ the oscillator strength
and $\xi$ the equilibrium inter-particle distance. The parameters $\mu$,
$\kappa$ and $\xi$ are intrinsic properties (the last two defining the
potential function). 

This kind of chain seems to be different from most that are studied in
literature: the positions of the chain particles are dynamical variables so
that the chain can move as a whole. However, the chain can still be solved by
methods that are described in \cite{Kittel,Rutherford}. 

First, we find the variables $u_n$ and $q_n$ that diagonalize the Hamiltonian
describing the so-called normal modes. The transformation is 
\begin{equation}\label{xu}
 x_n = \sum_{m=0}^{N-1}Y^m_nu_m + \left(n - \frac{N+1}{2}\right)\xi,
\end{equation}
and
\begin{equation}\label{pu}
p_n = \sum_{m=0}^{N-1}Y^m_nq_m,
\end{equation}
where the mode index $m$ runs through $0,1,\cdots,N-1$ and $Y^m_n$ is an
orthogonal matrix; for even $m$, 
\begin{equation}\label{evenm}
Y^m_n = A(m)\cos\left[\frac{\pi m}{N}\left(n-\frac{N+1}{2}\right)\right],
\end{equation}
while for odd $m$,
\begin{equation}\label{oddm}
Y^m_n = A(m)\sin\left[\frac{\pi m}{N}\left(n-\frac{N+1}{2}\right)\right]
\end{equation}
and the normalization factors are given by
\begin{equation}\label{factor}
A(0) = \frac{1}{\sqrt{N}},\quad A(m) = \sqrt{\frac{2}{N}},\quad m>0.
\end{equation}

To show that $u_n$ and $q_n$ do represent normal modes, we substitute Eqs.\
(\ref{xu}) and (\ref{pu}) into (\ref{Ham}) and obtain, after some calculation, 
$$
H = \frac{1}{2\mu}\sum_{m=0}^{N-1}q_m^2 + \frac{\mu}{2}\sum_{m=0}^{N-
  1}\omega_m^2u_m^2, 
$$
which is indeed diagonal. The mode frequencies are 
\begin{equation}\label{spectr}
\omega_m = \frac{2\kappa}{\sqrt{\mu}}\sin\frac{m}{N}\frac{\pi}{2}.
\end{equation}

Consider the terms with $m=0$. We have $\omega_0=0$, and
$Y^0_n=1/\sqrt{N}$. Hence, 
$$
u_0 = \sum_{n=1}^N\frac{1}{\sqrt{N}}x_n, \quad q_0 =
\sum_{n=1}^N\frac{1}{\sqrt{N}}p_n, 
$$
so that
$$
u_0 = \sqrt{N}X,\quad q_0 = \frac{1}{\sqrt{N}}P,
$$
where $X$ is the centre-of-mass coordinate of the chain and $P$ is its total
momentum. The 'zero' terms in the Hamiltonian then reduce to 
$$
\frac{1}{2M}P^2
$$
with $M = N\mu$ being the total mass. Thus, the 'zero mode' describes a
straight, uniform motion of the chain as a whole. The other modes are harmonic
oscillators called 'phonons' with eigenfrequencies $\omega_m$, $m =
1,2,\dots,N-1$. The spectrum of our system is built from the mode frequencies
by the formula 
\begin{equation}\label{phonons}
E = \sum_{m=1}^{N-1}\nu_m \hbar\omega_m,
\end{equation}
where $\{\nu_m\}$ is an $(N-1)$-tuple of non-negative integers---phonon
occupation numbers. 

At this stage, a new and important assumption must be done. We imagine that
all states $\rho$ of the modes $m = 1,\cdots,N-1$ are prepared that have the
same average internal energy $\bar{E}$, 
$$
\text{Tr}\left[\rho\left(H-\frac{P^2}{2M}\right)\right] = \bar{E}.
$$
We further assume that it is done in a perfectly random way, i.e., all other
conditions or bias are to be excluded. Hence, the resulting mixture must
maximize the entropy. In this way, the maximum of entropy does not represent
an additional condition but rather the absence of any. The resulting state
$\rho_{\bar{E}}$ is the Gibbs state of the internal degrees of freedom. The
conditions that define the preparation of Gibbs state are objective and need
not have to do with human laboratory activity. 

The internal energy has itself a very small relative variance in the Gibbs
state; this need not be assumed from the start. Thus, it is a classical
property. All other classical internal properties will turn out to be
functions of the classical internal energy. Hence, for the internal degrees of
freedom, $\bar{E}$ forms itself a complete set of state coordinates introduced
in Sec.\ 4. The mathematics associated with the maximum entropy principle is
variational calculus. The condition of fixed averaged energy is included with
the help of Lagrange multiplier denoted by $\lambda$. It becomes a function
$\lambda(\bar{E})$ for the resulting state. As it is well known,
$\lambda(\bar{E})$ has to do with temperature. 

The phonons of one species are excitation levels of a harmonic oscillator, so
we have 
$$
u_m = \sqrt{\frac{\hbar}{2\mu\omega_m}}(a_m + a^\dagger_m),
$$
where $a_m$ is the annihilation operator for the $m$-th species. The diagonal
matrix elements between the energy eigenstates $\mid\nu_m\rangle$ that we
shall need then are 
\begin{equation}\label{averu}
\langle\nu_m\mid u_m\mid\nu_m\rangle = 0,\quad \langle\nu_m\mid
u^2_m\mid\nu_m\rangle = \frac{\hbar}{2\mu\omega_m}(2\nu_m + 1). 
\end{equation}

For our system, the phonons of each species form statistically independent
subsystems, hence the average of an operator concerning only one species in
the Gibbs state $\rho_{\bar{E}}$ of the total system equals the average in the
Gibbs state for the one species. Such a Gibbs state operator for the $m$-th
species has the form 
$$
\rho_m = \sum_{\nu_m=0}^{\infty}\mid\nu_m\rangle
p_{\nu_m}^{(m)}\langle\nu_m\mid, 
$$
where
$$
p_{\nu_m}^{(m)} = Z^{-1}_{m}\exp\left(-\hbar\lambda\omega_m\nu_m\right)
$$
and $Z_m$ is the partition function for the $m$-th species
\begin{equation}\label{partf}
Z_{m}(\lambda) = \sum_{\nu_{m}=0}^\infty e^{-\lambda\hbar\omega_m\nu_m} =
\frac{1}{1-e^{-\lambda\hbar\omega_m}}. 
\end{equation}

\subsection{The length of the body}
The classical property that will be defined and calculated in our quantum
model is the average length of the body. Let us define the length operator by 
\begin{equation}\label{length}
  L = x_N - x_1.
\end{equation}
It can be expressed in terms of modes $u_m$ using Eq.~(\ref{xu}),
$$
L = (N-1)\xi + \sum_{m=0}^{N-1}(Y^m_N-Y^m_1)u_m.
$$
The differences on the right-hand side are non-zero only for odd values of
$m$, and equal then to $-2Y^m_1$. We easily find, using Eqs.~(\ref{oddm}) and
(\ref{factor}): 
\begin{equation}\label{L}
L = (N-1)\xi - \sqrt{\frac{8}{N}}\ \sum_{m=1}^{[N/2]}(-
1)^m\cos\left(\frac{2m-1}{N}\frac{\pi}{2}\right)\,u_{2m-1}. 
\end{equation}

The average length is obtained inserting Eq.\ (\ref{averu}),
\begin{equation}\label{averL}
\langle L\rangle_{\bar{E}} = (N-1)\xi.
\end{equation}
It is a function of intrinsic properties $N$, $\xi$ and $\bar{E}$.

Eq.\ (\ref{L}) is an important result. It shows that contributions to the
length are more or less evenly distributed over all odd modes. The even
distribution will lead to a very small variance of $L$ in Gibbs states. Let us
give the proof that the relative variance of the length is indeed small. The
proof is not trivial because the distribution is not constant. The relative
variance is 
$$
\frac{\Delta L}{\langle L\rangle_{\bar{E}}}= \frac{\sqrt{\langle
    L^2\rangle_{\bar{E}} - \langle L\rangle_{\bar{E}}^2}}{\langle
  L\rangle_{\bar{E}}}. 
$$
To estimate the variance $\Delta L$ to the leading order for large $N$, we
start with 
\begin{multline*}
\langle L^2\rangle_{\bar{E}} = (N-1)^2\xi^2 \\+
\frac{8}{N}\sum_{m=1}^{[N/2]}\sum_{n=1}^{[N/2]}(-1)^{m+n}
\cos\left(\frac{2m-1}{N}\frac{\pi}{2}\right)
\cos\left(\frac{2n-1}{N}\frac{\pi}{2}\right) \langle
u_{2m-1}u_{2n-1}\rangle_{\bar{E}}. 
\end{multline*}
Since
$$
\langle u_{2m-1}u_{2n-1}\rangle_{\bar{E}}=\delta_{mn}\langle
u_{2m-1}^2\rangle_{\bar{E}}, 
$$
the above formula leads to
$$
\langle L^2\rangle_{\bar{E}} - \langle L\rangle_{\bar{E}}^2 =
\frac{8}{N}\sum_{m=1}^{[N/2]} \cos^2\left(\frac{2m-1}{N}\frac{\pi}{2}\right)
\langle u_{2m-1}^2\rangle_{\bar{E}}, 
$$
where
$$
\langle u_{2m-1}^2\rangle_{\bar{E}} = \frac{1}{Z_{2m-1}}
\sum_{\nu_{2m-1}=0}^\infty \frac{\hbar}{2\mu\omega_{2m-1}}(2\nu_{2m-1}+1)
\exp(-\lambda\hbar\omega_{2m-1}\nu_{2m-1}). 
$$
Introducing dimensionless quantities
$$
x_m = \sin\left(\frac{2m-1}{N}\frac{\pi}{2}\right),\quad \gamma =
\frac{2\hbar\kappa\lambda}{\sqrt{\mu}}, 
$$
we can substitute $\omega_{2m-1}=(2 \kappa/\sqrt{\mu})x_m$ and obtain the
intermediate result 
$$
\langle L^2\rangle_{\bar{E}} - \langle L\rangle_{\bar{E}}^2 =
\frac{2}{N}\frac{\hbar}{\kappa\sqrt{\mu}}
\sum_{m=1}^{[N/2]}\frac{1-x_m^2}{x_m} \frac{1+e^{-\gamma x_m}}{1-e^{-\gamma
    x_m}}. 
$$
In order to extract the leading term for large $N$, we note that
$$
x_m-x_{m-1} = \frac{\pi}{N}\cos\frac{2m-1}{N}\frac{\pi}{2} + O(N^{-2}).  
$$
Then we can write
$$
\langle L^2\rangle_{\bar{E}} - \langle L\rangle_{\bar{E}}^2 \approx
\frac{2}{\pi}\frac{\hbar}{\kappa\sqrt{\mu}}\sum_{m=1}^{[N/2]}(x_m-x_{m-
  1})f(x_m), 
$$
where
$$
f(x)= \frac{\sqrt{1-x^2}}{x}\ \frac{1+e^{-\gamma x}}{1-e^{-\gamma x}}.
$$
By inspection, $f$ is a decreasing function od $x$ in the interval $(0,1)$
diverging to plus infinity at $x\rightarrow 0+$ and going through zero at
$x=1$. The leading term at $x\rightarrow 0+$ is 
$$
f(x) = \frac{2}{\gamma x^2}[1+O(x)].
$$
The block diagram of the sum now shows that
$$
\sum_{m=1}^{[N/2]}(x_m-x_{m-1})f(x_m) < 2x_1f(x_1) + \int_{x_1}^1dx\,f(x).
$$
The dependence of the integral on its lower bound can be approximated by
$$
\int_{x_1}^1dx\,f(x) = \text{const} + \frac{2}{\gamma x_1}[1+O(x_1)].
$$
Thus, the leading term in the sum is $ 6/\gamma x_1 \approx 12N/\gamma\pi$ so
that the leading term in $\langle L^2\rangle_{\bar{E}} - \langle
L\rangle_{\bar{E}}^2 $ is $ (12/\pi^2\lambda\kappa^2)N$. We obtain the final
result valid for large $N$ 
\begin{equation}
\frac{\Delta L}{\langle L\rangle_{\bar{E}}} \approx
\frac{2\sqrt{3}}{\pi\kappa\xi\sqrt{\lambda}}\frac{1}{\sqrt{N}}. 
\end{equation}

Thus, the small relative variance for large $N$ need not be assumed from the
start. The only assumptions are values of some structural properties and that
an average value of energy is fixed. In the sense explained in Section 4, the
length is then a classical property of our model body. We have obtained even
more information: the internal-energy independence of the length (in this
model, the dependence is trivial). This is an objective relation that can be
in principle tested by measurements. 

Similar results can be obtained for further thermodynamic properties such as
elasticity coefficient, specific heat etc. They all are well known to have
small variances in Gibbs states. The reason is that the contributions to these
quantities are homogeneously distributed over the normal modes and the modes
are mechanically and statistically independent. Further important classical
properties are the mechanical ones: centre of mass and total momentum. In
fact, these quantities can be chosen as the rest of the state coordinates for
the whole chain. The contributions to them are perfectly homogeneously
distributed over all atoms, not modes: the bulk motion is mechanically and
statistically independent of all other modes and so its variances will not be
small in Gibbs states. Still, generalized statistical methods can be applied
to it. This is done in a separate paper \cite{hajicek}. 

The last remark is that the thermodynamic equilibrium can settle down starting
from an arbitrary state only if some weak but non-zero interaction exists
between the phonons. This can easily be arranged so that the influence of the
interaction on our result is negligible.

\subsection*{Acknowledgements}
P.H. is indebted to Juerg Gasser and Uwe-Jens Wiese for reading the first
version of the manu\-script and suggesting many improvements in the
text. Important remarks have been sent by Jim Hartle. Anonymous referees have
also contributed by their criticisms to better understanding. Thanks go to the
Institute of Theoretical Physics, Faculty of Mathematics and Physics of the
Charles University, as well as the Doppler Seminar of the Department of
Physics, Faculty of Nuclear Sciences and Physical Engineering, Czech Technical
University Prague for hospitality and discussion. J.T. gratefully acknowledges
partial support by the Ministry of Education of Czech Republic (projects
MSM6840770039 and LC06002).


\begin{thebibliography}{99}
\bibitem{Griffiths}R.~B.~Griffiths, {\it Consistent Quantum Theory}, Cambridge
  University Press, Cambridge, UK, 2002.   
\bibitem{Nikolic}H.~Nikolic, Quantum mechanics: myths and facts. ArXiv:
  quant-ph/0609163.   
\bibitem{Isham}C.~J.~Isham, {\it Lectures on Quantum Theory. Mathematical and
    Structural Foundations}, Imperial College Press, London 1995. 
\bibitem{d'Espagnat}B. d'Espagnat, {\em Veiled Reality}, Addison-Wesley,
  Reading, 1995.  
\bibitem{Peres}A.~Peres, {\it Quantum Theory: Concepts and Methods}, Kluwer,
  Dordrecht, 1995. 
\bibitem{ludwig}G. Ludwig, {\it Foundations of Quantum Mechanics}, Springer,
  Berlin, 1983.   
\bibitem{kraus}K. Kraus, {\it States, Effects, and Operations}, Springer
  Lecture Notes in Physics 190, Berlin, 1983. 
\bibitem{Bub}J.~Bub, {\it Interpreting the Quantum World}. Cambridge
  University Press, Cambridge (UK), 1997. 
\bibitem{Dirac}P.~A.~M.~Dirac, {\it Quantum Mechanics}, 4th edition. Oxford,
  Clarendon Press, 1958. 
\bibitem{JvN}J.~von~Neumann, {\it Mathematical Foundation of Quantum
    Mechanics}, Princeton University Press, Princeton NJ, 1955. 
\bibitem{TH}J.~Tolar and P.~H\'aj\'{\i}\v{c}ek, Phys. Letters A {\bf 353}
  (2006) 19. 
\bibitem{Schroeck}F.~E.~Schroeck, {\it Quantum Mechanics on Phase Space}
\bibitem{hajicek}P.~H\'aj\'{\i}\v{c}ek, Quantum model of classical mechanics:
  maximum entropy packets. ArXiv:0901.0436. 
\bibitem{Piron}C.~Piron, {\it Foundations of Quantum Physics}, Benjamin,
  Reading 1976; C.~Piron, Found.\ Phys.\ {\bf 2} (1972) 287. 
\bibitem{BvN}G.~Birkhoff and J.~von~Neumann, Ann. of Math. {\bf 37} (1936) 823.
\bibitem{thirring}W. Thirring, {\it Lehrbuch der Mathematischen Physik},
  Springer, Berlin, 1980. 
\bibitem{Leggett}A.~J.~Leggett, J. Phys.: Cond.\ Mat.\ {\bf 14} (2002) R415.
\bibitem{zeh}D.~Giulini, E.~Joos, C.~Kiefer, J.~Kupsch, I.-O.~Stamatescu,
  H.~D.~Zeh, {\it Decoherence and the Appearance of Classical World in Quantum
    Theory}, Springer, Berlin, 1996. 
\bibitem{Exner}F.~Exner, {\it Vorlesungen \"{u}ber die physikalischen
    Grundlagen der Naturwissenschaften}, Deuticke, Leipzig, 1922. 
\bibitem{Born}M.~Born, Phys.~Bl\"{a}tter {\bf 11} (1955) 49.
\bibitem{Jaynes}E.~T.~Jaynes, {\it Probability Theory. The Logic of Science},
  Cambridge University Press, Cambridge (UK) 2003. 
\bibitem{kampen}N.~G.~van Kampen, Physica A {\bf 194} (1993) 542.
\bibitem{zurek}W.~H.~Zurek, Rev.\ Mod.\ Phys.,{\bf 75} (2003) 715.
\bibitem{poulin} D.~Poulin, Phys.\ Rev.\ A {\bf 71} (2005) 022102.
\bibitem{Kofler}J.~Kofler and \v{C}.~Brukner, Phys.\ Rev.\ Lett. {\bf 99}
  (2007) 180403. 
\bibitem{Hepp}K.~Hepp, Helvetica Phys.~Acta, {\bf 45} (1972) 237.
\bibitem{Bell3}J.~S.~Bell, Helv. Phys.~Acta, {\bf 48} (1975) 93.
\bibitem{Bona}P.~B\'ona, Acta Phys.~Slov., {\bf 23} (1973) 149, {\bf 25}
  (1975) 3, {\bf 27} (1977) 101. 
\bibitem{Sewell}G.~L.~Sewell, {\it Quantum Mechanics and its Emergent
    Macrophysics}, Princeton University Press, Princeton, 2002. 
\bibitem{bell2}J. S. Bell, Against 'measurement' in {\it Sixty Two Years of
    Uncertainty}, A. I. Miller (Ed.), Plenum, New York, 1990. 
\bibitem{Wallace}D.~Wallace, The quantum measurement problem: state of play,
  arXiv:0712.0149v1 [quant-ph]. 
\bibitem{Jauch}J.~M.~Jauch, Helv. Phys. Acta {\bf 37} (1964) 293.
\bibitem{Kittel}C.~Kittel, {\it Introduction to Solid State Physics}, Wiley,
  New York 1976. 
\bibitem{Rutherford} D.~E.~Rutherford, Proc.\ Roy.\ Soc.\ (Edinburgh), Ser.~A,
  {\bf 62} (1947), 229; {\bf 63} (1951), 232. 
\end{thebibliography}
\end{document}